\documentclass[iop]{emulateapj}
\usepackage{apjfonts}
\usepackage{graphicx}
\usepackage{color}
\usepackage{amsmath}

\def\d3{$\delta_{3}$ }
\def\1d3{$(1 + \delta_{3})$ }
\def\l1d3{$\log_{10}(1 + \delta_{3})$ }

\def\s3{$\Sigma_{3}$}
\def\ha{H$\alpha$}
\def\hb{H$\beta$}

\def\othree{[OIII] 5007}
\def\ntwo{[NII] 6584}

\def\24m{24 $\mu$m}

\def\sm{$\rm~M_{*}$}

\def\kms{${\rm km~s^{-1}}$ }

\def\Msolar{$\rm M_{\odot}$}

\def\rmxaa{RMxAA}

\def\sigsm{$\Sigma_{*}$}
\def\sigsfr{$\Sigma_{\rm~SFR}$}

\def\h2{$\rm~H_{2}$}
\def\Mh2{$\rm~M_{H2}$}
\def\sigh2{$\Sigma_{\rm~H_{2}}$}
\def\fgas{$f_{\rm~gas}$}

\def\co{$^{12}$CO(1-0)}

\shorttitle{Scaling relations of nearby galaxies}
\shortauthors{Lin et al.}

\begin{document}

\title{The ALMaQUEST Survey: The molecular gas main sequence and the origin of the star forming main sequence}

\author{Lihwai Lin \altaffilmark{1}, Hsi-An Pan \altaffilmark{1}, Sara L. Ellison \altaffilmark{2},  Francesco Belfiore \altaffilmark{3}, Yong Shi \altaffilmark{4,5},
 Sebasti\'{a}n F. S\'{a}nchez \altaffilmark{6}, Bau-Ching Hsieh \altaffilmark{1}, Kate Rowlands \altaffilmark{7}, S. Ramya \altaffilmark{8}, 
 Mallory D. Thorp \altaffilmark{2}, Cheng Li \altaffilmark{9}, Roberto Maiolino \altaffilmark{10,11}}

\altaffiltext{1}{Institute of Astronomy \& Astrophysics, Academia Sinica, Taipei 10617, Taiwan; Email: lihwailin@asiaa.sinica.edu.tw}
\altaffiltext{2}{Department of Physics \& Astronomy, University of Victoria, Finnerty Road, Victoria, British Columbia, V8P 1A1, Canada}
\altaffiltext{3}{European Southern Observatory, Karl-Schwarzschild-Str. 2, Garching bei M$\ddot{u}$nchen, 85748, Germany}
\altaffiltext{4} {School of Astronomy and Space Science, Nanjing University, Nanjing 210093, People's Republic of China}
\altaffiltext{5} {Key Laboratory of Modern Astronomy and Astrophysics (Nanjing University), Ministry of Education, Nanjing 210093, People's Republic of China}
\altaffiltext{6}{Instituto de Astronom\'ia, Universidad Nacional Aut\'onoma de  M\'exico, Circuito Exterior, Ciudad Universitaria, Ciudad de M\'exico 04510, Mexico}
\altaffiltext{7} {Department of Physics \& Astronomy, Johns Hopkins University, Bloomberg centre, 3400 N. Charles St., Baltimore, MD 21218, USA}
\altaffiltext{8} {Indian Institute of Astrophysics, II Block, Koramangala, Bengaluru 560 034, INDIA}
\altaffiltext{9} {Tsinghua Center for Astrophysics and Physics Department, Tsinghua University, Beijing 100084, China}
\altaffiltext{10}{Cavendish Laboratory, University of Cambridge, 19 J. J. Thomson Avenue, Cambridge CB3 0HE, United Kingdom}
\altaffiltext{11}{University of Cambridge, Kavli Institute for Cosmology, Cambridge, CB3 0HE, UK.}

\begin{abstract}
The origin of the star forming main sequence ( i.e., the relation between star formation rate and stellar mass, globally or on kpc-scales; hereafter SFMS) remains a hotly debated topic in galaxy evolution. Using the ALMA-MaNGA QUEnching and STar formation (ALMaQUEST) survey, we show that for star forming spaxels in the main sequence galaxies, the three local quantities, star-formation rate surface density (\sigsfr), stellar mass surface density (\sigsm), and the \h2~mass surface density (\sigh2), are strongly correlated with one another and form a 3D linear (in log) relation with dispersion. In addition to the two well known scaling relations, the resolved SFMS (\sigsfr~ vs. \sigsm) and the Schmidt-Kennicutt relation (\sigsfr~ vs. \sigh2; SK relation), there is a third scaling relation between \sigh2~ and \sigsm, which we refer to as the `molecular gas main sequence' (MGMS). The latter indicates that either the local gas mass traces the gravitational potential set by the local stellar mass or both quantities follow the underlying total mass distributions. The scatter of the resolved SFMS ($\sigma \sim 0.25$ dex) is the largest  compared to those of the SK and MGMS relations ($\sigma \sim$ 0.2 dex). A Pearson correlation test also indicates that the SK and MGMS relations are more strongly correlated than the resolved SFMS. Our result suggests a scenario in which the resolved SFMS is the least physically fundamental and is the consequence of the combination of the SK and the MGMS relations.

\end{abstract}


\section{INTRODUCTION}

The discovery of the tight relation between the star formation rate and the stellar mass of galaxies, namely, the `star forming main sequence' \citep[SFMS or MS;][]{bri04,noe07,dad07,lin12,whi12,spe14}, not only offers a channel to characterize properties of galaxies but also provides constraints on the galaxy formation and evolution models. However, the physics driving this scaling relation are not well understood, as it is not clear why the current star formation rate is related to the total star formation rate integrated over the past (i.e., stellar mass). Star formation is a complex process that involves multiple scales. For example, whereas the global star formation rate depends on the large-scale environment \citep{dre80,kau04,coo07,elb07,lin14}, the efficiency of gas converted into stars is dependent on local conditions operating on sub-kpc scales \citep{kru05,mur11}. Probing relationships between stars and gas across different physical scales may therefore shed light on the origin of the star-forming main sequence.

Recent studies using Integral Field Spectrosocpy (IFS) observations have shown that the star-formation rate surface density (\sigsfr) traces the stellar mass surface density (\sigsm) linearly at kpc/sub-kpc scales \citep{san13,wuy13,can16,hsi17,pan18,ell18,med18,vul19}. This so-called `resolved' SFMS (hereafter rSFMS) indicates that the connection between the global star formation rate and stellar mass may actually originate from  local processes. 
However, whilst the relationship between \sigh2~ and \sigsfr~ \citep[Schmidt-Kennicutt or SK relation;][]{sch59,ken98} is understood as the formation of stars from molecular gas, the physical reason for the rSFMS remains a mystery.
To complete the picture of the origin of rSFMS, it is therefore vital to relate the molecular gas to star formation tracers and stellar masses with the same spatial resolution. In this work, we combine spatially resolved observations from the Mapping Nearby Galaxies at Apache Point Observatory \citep[MaNGA;][]{bun15} and Atacama Large
Millimeter Array (ALMA)  for 14 star-forming main sequence (MS) galaxies at $z \sim 0.03$, which allow us to study the relationships between the surface densities of star formation rate, molecular gas and stellar mass on kpc scales.

Throughout this paper we adopt the following cosmology: \textit{H}$_0$ = 70~\kms Mpc$^{-1}$, $\Omega_{\rm m} = 0.3$ and $\Omega_{\Lambda } = 0.7$. We use a Salpeter IMF.

\begin{deluxetable}{lccc}
\tabletypesize{\scriptsize}
\tablewidth{0pt}
\tablecaption{Best-fit parameteres ($a$ and $b$) and associated scatters ($\sigma$) for the 2-D scaling relations\label{tab:fit}}
\tablehead{
    \colhead{Relation} &
    \colhead{$a$ (ODR)} &
    \colhead{$b$ (ODR)} &
     \colhead{$\sigma$ (ODR)} 
    }
\startdata

\enddata
rSFMS (\sigsfr vs. \sigsm) & $1.19 \pm 0.01$ & $-11.68 \pm 0.11$ & 0.25 \\
 SK (\sigsfr vs. \sigh2) & $1.05 \pm 0.01$ & $-9.33 \pm 0.06$ & 0.19 \\
 MGMS (\sigh2 vs. \sigsm) & $1.10 \pm 0.01$ & $-1.95 \pm 0.08$ & 0.20
\end{deluxetable}

\begin{figure*}
\centering
\includegraphics[angle=0,width=0.9\textwidth]{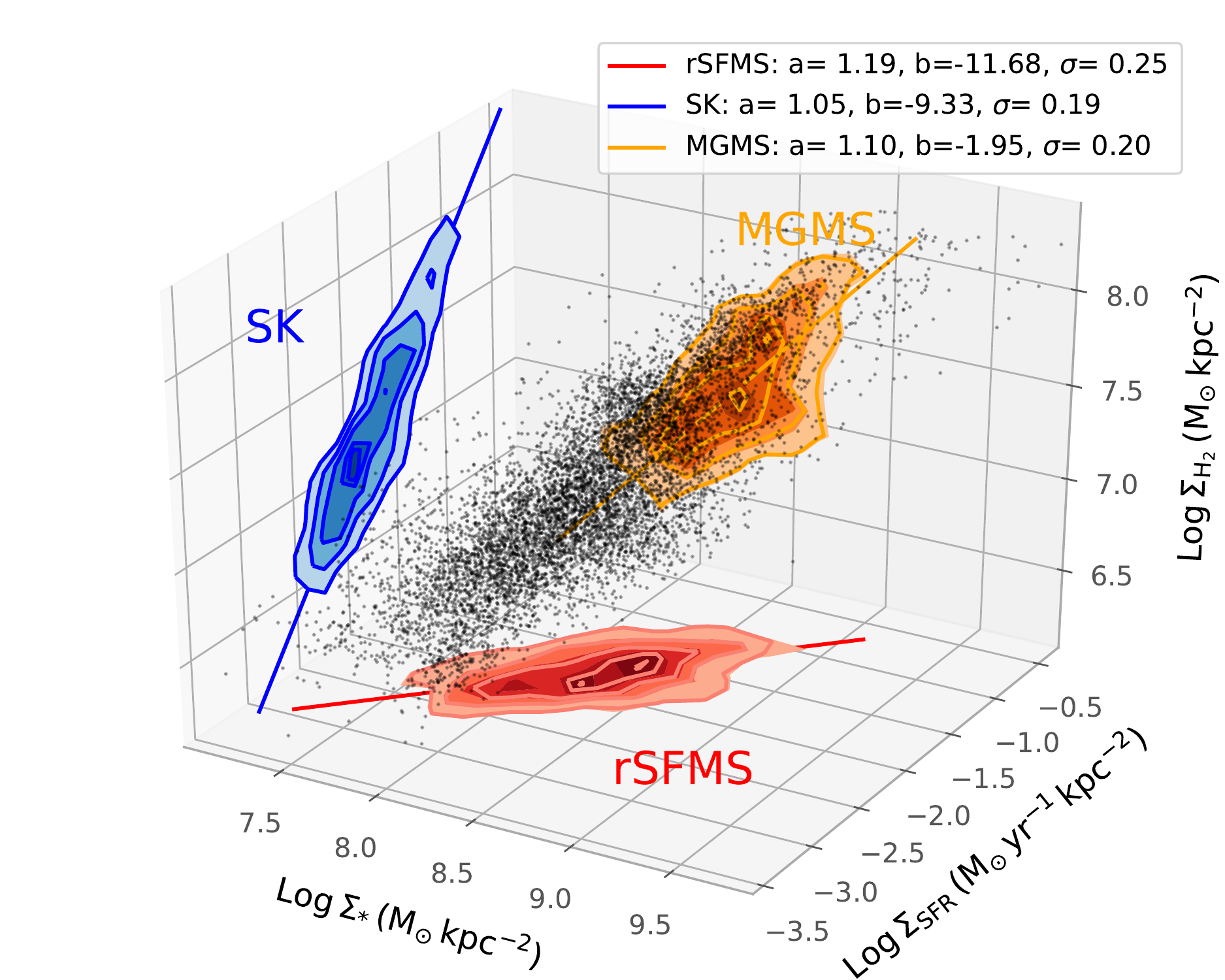}
\caption{The 3D distribution between \sigsm, \sigh2, and \sigsfr, computed for 5383 spaxels (black points) identified as star-forming regions taken from 14 MaNGA main sequence galaxies. The contours show the results projected on the 2D planes (red: SFMS; blue: SK; orange: MGMS) , with the contour levels corresponding to 20\%, 40\%, 60\%, 80\%, and 90\% of the density peaks. The best-fit parameters and associated scatters ($\sigma$) based on ODR fitting are given in the legend.  \label{fig:3d}}
\end{figure*}

\section{SAMPLE and OBSERVATIONS \label{sec:data}}

The ALMA-MaNGA QUEnching and STar formation survey (ALMaQUEST; L. Lin et al. in prep.) is a compilation of four ALMA PI programs that follow up MaNGA galaxies with \co~at a spatial resolution matched to MaNGA (FWHM $\sim 2.5 \arcsec$). The `Quenching' component (2015.1.01225.S, 2017.1.01093.S \& 2018.1.00558.S; PI:L. Lin) of ALMaQUEST targets 32 galaxies that are on the main sequence ($\sim$1/3 of the sample) and those in the green valley ($\sim$ 2/3 of the sample). The other component, `Starburst' program (2018.1.00541.S; PI: S. Ellison), consists of 12 central starburst galaxies and 4 regular main sequence galaxies (see S. Ellison et al., in prep.).  All of these observations adopt identical observing setups and reduction procedures. 
In this work, we present results using 14 MS galaxies with $10 <$ log(\sm/\Msolar) $< 11.5$ taken from the ALMaQUEST survey. These galaxies are selected to have  $10^{-10.5}$ yr$^{-1} <$ specific star formation rate (sSFR)  $<10^{-9.5}$ yr$^{-1} $ without showing strong central starburst features. The sSFR range is sufficiently broad to ensure that we sample a variety of star-forming galaxies.
 The CO data is processed following the procedures described in \citet{lin17} and the details will be described in the ALMaQUEST survey paper (L. Lin et al., in prep.). The H$_{2}$ mass surface density (\sigh2) is computed from the inclination-corrected CO surface density by adopting a conversion factor ($\alpha_{\mathrm{CO}}$) of 4.3 \Msolar (K km s$^{-1}$ pc$^{2}$)$^{-1}$ \citep[e.g.,][]{bol13}. An S/N $> 2$ cut \footnote{Adopting different S/N cuts between 1.5 to 3 do not significantly alter the slopes of the scaling relations presented here and none of our conclusions are affected. We choose to adopt a loose cut in CO in order to maximize the number of spaxels that can be used in this work.} in the CO line is applied to our analysis.

Other measurements, such as \sigsm~ and emission-line fluxes,  are obtained from the MaNGA Data Release 15 (DR15) data cubes processed by the Pipe3D pipeline \citep{san16a}. All the emission lines were then dust extinction corrected using the Balmer decrement computed at each spaxel, following the method described in the Appendix of \citet{vog13}. An extinction law with Rv = 4.5 \citep{fis05} is used. The SFR is estimated based on this extinction corrected \ha~ flux using the
conversion given by Kennicutt (1998) with the Salpeter IMF. \sigsm~ and \sigsfr~ are computed using the stellar mass and SFR derived for each spaxel, normalized to the physical area of one spaxel with the inclination correction applied. 
We limit our sample to spaxels with \sigsm $> 10^{7}$ \Msolar~ kpc$^{-2}$ and require that the S/N in strong lines (\ha~and \hb) and in weak lines (\ntwo~and \othree) to be $>$ 3 and $>$ 2, respectively. These spaxels are further classified as star-forming regions using the \ntwo~diagnostic, following the method of \citet{kau03}.

\begin{figure}
\centering
\includegraphics[angle=0,width=0.5\textwidth]{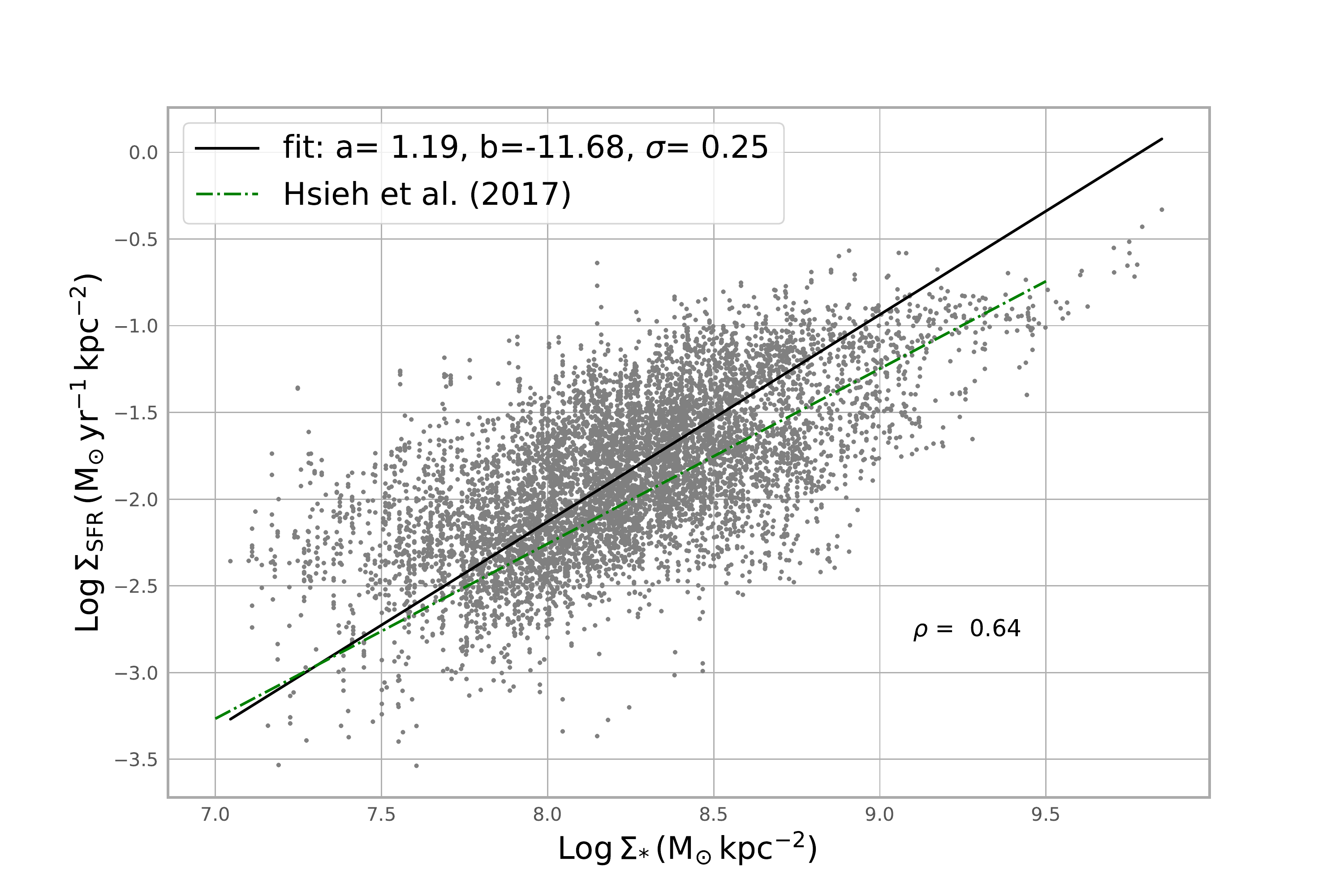}
\caption{The spaxel-based star-formation rate surface density vs. stellar mass surface density (\sigsfr~--\sigsm) relation (grey points) for the ALMaQUEST sample. The black solid line represents the best fit to our data. The best-fit parameters, associated scatter ($\sigma$), and the Pearson correlation coefficient ($\rho$) are given in the legend. The green dotted dashed line is the ODR fitting result derived by \citet{hsi17} based on 536 star forming main sequence galaxies in the MaNGA DR13 sample. \label{fig:sfrsm}}
\end{figure}

\begin{figure*}
\centering
\includegraphics[angle=0,width=0.9\textwidth]{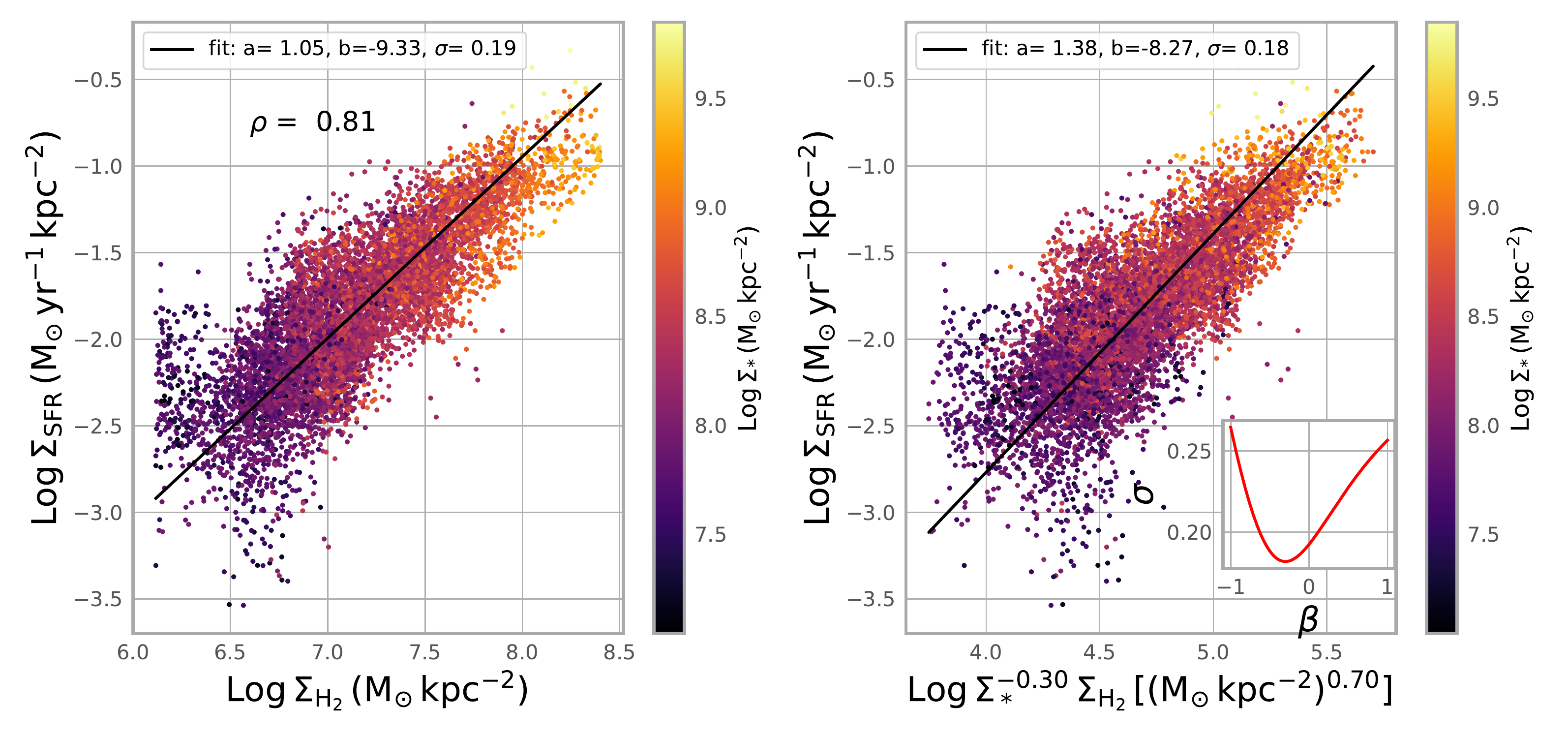}
\caption{The spaxel-based Schmidt-Kennicutt (left panel) and extended Schmidt-Kennicutt (right panel) relations in our sample. The black solid lines show the best fits to our data. The best-fit parameters, associated scatters ($\sigma$), and the Pearson correlation coefficient ($\rho$) are given in the legend. The inset of the right panel shows the associated scatter of the extended SK relation as a function of the power exponent $\beta$. \label{fig:sfrh2}}
\end{figure*}

\begin{figure}
\centering
\includegraphics[angle=0,width=0.5\textwidth]{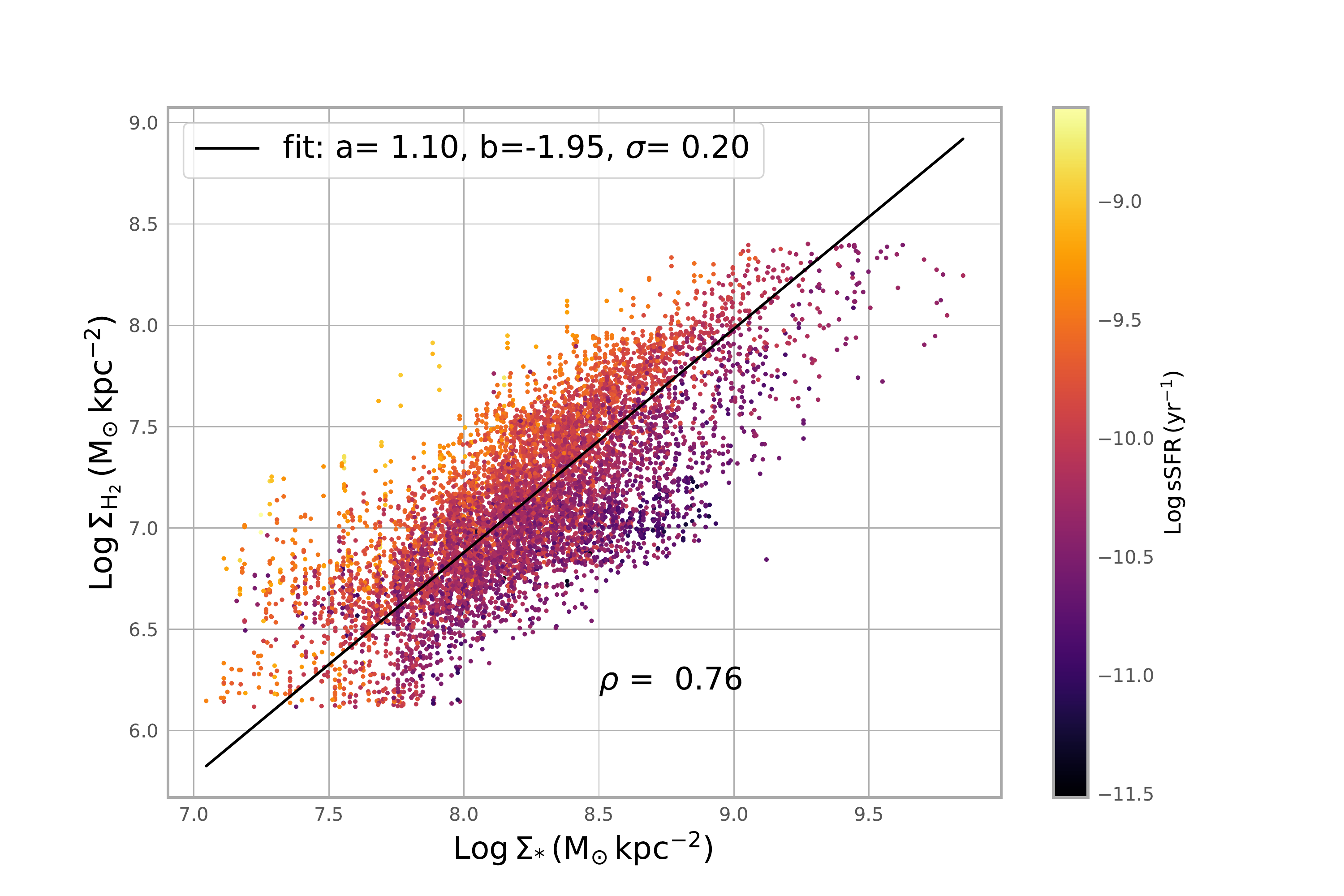}
\caption{The spaxel-based molecular gas main sequence (\sigh2--\sigsm) relation, color-coded by sSFR. The black solid line shows the best fit to our data. The best-fit parameters, associated scatter ($\sigma)$, and the Pearson correlation coefficient ($\rho$) are given in the legend.  \label{fig:h2sm}}
\end{figure}

\section{RESULTS}

\subsection{The 3D and 2D scaling relations between \sigsfr, \sigh2, and \sigsm}
We first made a 3D plot (Figure \ref{fig:3d}) to show the spaxel-to-spaxel relationship between \sigsfr, \sigsm, and \sigh2~ for star-forming spaxels identified using the diagnostic described in Section 2. Figure \ref{fig:3d} shows that these three quantities form a 3D linear (in log) relation with dispersion, suggesting that each pair of these three variables forms a tight relation. This is further illustrated by the three contours that represent the de-projected data points on the \sigsfr--\sigsm (red), \sigsfr--\sigh2 (blue), and \sigh2--\sigsm~ (orange) planes. In addition to the well known SK relation (\sigsfr~ vs. \sigh2) and resolved SFMS ( \sigsfr~ vs. \sigsm), we also find that \sigh2~ traces \sigsm, which we hereafter refer to as the 'molecular gas main sequence' (MGMS). We fit each of the above relations using the orthogonal distance regression (ODR) fitting method with a power law parametrized as the following:

\begin{align}\label{eq: 2d}
log_{10} \Sigma_{SFR} &= a*log_{10} \Sigma_{*} + b\\
log_{10} \Sigma_{SFR} &= a*log_{10} \Sigma_{H2} + b\\
log_{10} \Sigma_{H2} &= a*log_{10} \Sigma_{*} + b
\end{align}

The best-fit parameters are shown in the legends of Figure \ref{fig:3d} and in Table \ref{tab:fit}.

\subsection{Resolved Star-forming Main Sequence (rSFMS)}

Figure \ref{fig:sfrsm} shows \sigsfr~vs. \sigsm~ relation for the star-forming spaxels in our sample.
We first note that the slope of the resolved main sequence (grey points), $1.19\pm0.01$, is higher than those ($\leq 1$) reported in the literature \citep{can16,hsi17,pan18,ell18,med18,can19}. 
It is already known that the slope is sensitive to both the fitting algorithms (e.g.,  ODR vs. ordinary least squares) and whether the non-HII regions are excluded or not. None the less, our result is still slightly than that reported ($\sim 1$) by \citet{hsi17} who computed \sigsfr~and \sigsm~ using the same method for 536 star-forming galaxies from the MaNGA DR13 sample with the ODR fitting. 
To test whether this is due to the limited number of galaxies used in this analysis, we randomly select 14 galaxies from the MaNGA DR13 star-forming population to measure the slope of the rSFMS and repeat this process 1000 times. The derived mean slope of the 1000 trials is 1.11$\pm$0.16. The slope of our CO sample is therefore consistent ($\sim$0.5$\sigma$) with the Monte-Carlo result, implying that steeper slope obtained for our sample is likely due to the small number of the CO targets instead of a biased population.

\subsection{SK and Extended SK Relation}

The SK relation has been measured to have a power law index ($N$) ranging from  0.5 to 3 \citep[see][]{big08}, depending on not only the star formation rate or gas tracer \citep{big08,gao04} but also the physical scale used when computing the surface density \citep{ono10,kre18}. On larger scales (e.g., averaged over the entire galaxy), $N$ is superlinear ($\sim$1.4) for \h2~ tracers, such as CO, and is close to unity for dense gas tracers, such as HCN or HCO+ \citep[e.g.,][]{gao04}. On the other hand, at smaller scales (kpc or sub-kpc), $N$ $\sim$ 1 or even lower for CO-based \h2~mass \citep{big08,rah16,bol17,kre18}. In this work, our data is well fitted by a power law with exponent $N \sim 1.05\pm0.01$ (the left panel of Figure \ref{fig:sfrh2}), in good agreement with recent studies on kpc/sub-kpc scales \citep{big08,ler13,bol17,kre18,dey19}. It is worth noting that the slope of the SK relation is not affected by the cut off in the \h2~limit associated with the S/N cut in the CO flux. This is because the number density of spaxels in that regime is relatively sparse.

Next, we explore the so-called extended SK relation \citep{shi11,shi18}, in which SFR is parametrized as SFR $\propto$ (\sigh2 $\times$ \sigsm$^{\beta}$)$^{a}$. It has been suggested that the scatter of the extended SK relation can be reduced when adopting $\beta$ = 0.5, which is often attributed to the effect of the mid-plane pressure \citep{ost10,hug13,shi11}. To test whether the extended SK relation applies to our sample,  we first color code our data with \sigsm~ in the left panel of Figure \ref{fig:sfrh2}.  It can be seen that there is no  apparent dependence of the scatter on \sigsm, contrary to expectations from an extended SK model.
On the other hand, a tendency of increasing \h2~with increasing \sigsm~is revealed, which will be discussed in \ref{sec:mgms}. To further explore the possibility of a \sigsm~ component to the SK relation, we vary the exponent $b$ between -1 and 1 and compute the scatter of the best-fit for a given $b$. In the inset of the right panel of Figure \ref{fig:sfrh2}, we plot the residual scatter against the  power exponent $b$. It is found that the scatter of the SK relation reaches a minimum value at $\beta = -0.3$ but is not significantly different from the case in the original SK relation (i.e., $\beta = 0$). The scatter even becomes larger when adopting the 'canonical' value of 0.5.

The extended SK relation with the optimal power exponent ($b$ = -0.3) is shown in the right panel of Figure \ref{fig:sfrh2}. Overall, across the range in the stellar mass surface density of our data, we do not find a significant improvement as seen in \citet{shi18} when adopting the extended SK relation. However, we note that our sample spans only 1.5 orders of magnitude in \sigsm~down to $10^{7}$ \Msolar~kpc$^{-2}$ while \citet{shi18} covers a much wider range in \sigsm~($\sim$ 5 orders of magnitude).

\subsection{Molecular Gas Main Sequence (MGMS)}\label{sec:mgms}
Unlike the rSFMS and SK relations, the relationship between \sigh2~and \sigsm~ has not been explored much in the literature. Using the gas surface density inferred from the Balmer decrement, \citet{bar18} found a weak correlation between gas surface density and stellar mass surface density with a large spread. In fact,  a positive correlation between these two quantities may be expected within individual galaxies, since both the gas and stellar mass profiles generally decline with radius in spiral galaxies \citep[e.g.,][]{cas17}. However, it is not clear whether there exists a universal scaling relation applicable to all systems. 

In Figure \ref{fig:h2sm}, we explore the spaxel-based correlation between these two quantities, color-coded by sSFR. We see that \sigh2~tracers is almost linearly dependent on \sigsm with a scatter of $\sim$ 0.2 dex, forming the MGMS. Again the \h2~limit has little impact on the derived slope of the MGMS in our data. It is found that spaxels with higher sSFR tend to lie on the upper end of the MGMS, meaning that the star formation is boosted in regions with enhanced gas fraction, as seen in spatially unresolved data \citep{sai17}. The effect of the gas fraction on sSFR locally will be further discussed in a companion ALMaQUEST paper \citep{ell19}.

\section{Discussion}
In this work, we have established a 3-way scaling relationship between \sigsfr, \sigsm, and \sigh2~on kpc scales. Each pair of these three parameters exhibits a tight correlation with a scatter of $\sim$0.19 -- 0.25 dex. Among the three relations, two already well known: SK relation (\sigsfr~ vs. \sigh2) and the rSFMS (\sigsfr~ vs. \sigsm), while the third one, the MGMS between \sigh2~ and \sigsm, is shown convincingly here for the first time. It is natural to ask: which of these correlations are more fundamental? 

To quantify the relative importance among the three 2D relations, we compute Pearson correlation coefficients ($\rho$) for each of them (shown in Figures \ref{fig:sfrsm} -- \ref{fig:h2sm}). This analysis shows that the SK relation has the strongest correlation, followed by the MGMS and then the rSFMS. Indeed, the SK relation is physically the most intuitive as stars form directly from the molecular clouds. It also has the smallest scatter of the three. On the other hand, the physical reason behind the MGMS  is less obvious. We consider two possible explanations for the presence of this correlation.  In the first scenario, the contribution of dark matter to the gravitational potential of the disk is negligible compared to baryonic components. Therefore, the local potential well of the disk is primarily set by the local \sigsm~ given that  the surface molecular gas fraction \fgas~ (defined as \sigh2/(\sigh2 + \sigsm)) in our sample is on the order of 10\% only. As a consequence, the gas follows the distribution of stellar mass, leading to the \sigh2~--\sigsm~linear correlation. In this scenario, the correlation between \sigh2~and \sigsm~might be expected to break down at higher redshifts where gas masses can exceed stellar masses  \citep{pop14,tac18,isb18}. Alternatively, if the dark matter dominates the mass distributions, both stars and gas will respond to the same gravity. In this case, the correlation between \sigh2~ and \sigsm~is caused by the underlying gravitational potential. Therefore, the \sigh2~--\sigsm~relation will still hold regardless of gas fraction. Dynamical measurements and the studies of the \sigh2~--\sigsm~relation as a function of cosmic time will shed light on the origin of the gas and stellar mass correlation.

Having established the MGMS, here we provide a plausible explanation to the empirical rSFMS.  If \sigsfr~$\propto$ \sigh2$^{a}$ and \sigh2~ $\propto$ \sigsm$^{b}$, one would expect \sigsfr~ $\propto$ \sigsm$^{a*b}$ = \sigsm$^{c}$. In our case, $c$ is measured to be 1.19, close to 1.16, the product of $a$ (1.05) and $b$ (1.10). Among the three relations, the rSFMS has the largest scatter ($\sigma$ = 0.25) and is close to the square root of the quadratic sum of the scatters ($\sigma$ = 0.28) from the SK (0.19) and MGMS (0.20) relations. Furthermore, the rSFR also has the smallest Pearson correlation coefficient. All these suggest that rSFMS could be a natural consequence of the other two relations. Finally, we note that our analyses presented are restricted to the star-forming spaxels of MS galaxies. The scaling relations of the retired spaxels and for galaxies deviated from the MS will be further explored in future works (Lin et al. in prep.)

\section{SUMMARY}\label{sec:summary}

Combining the ALMA \co~ and MaNGA observations of 14 main sequence galaxies taken from the ALMaQUEST survey, we investigate the relationships between the surface densities of star formation rate, molecular gas, and stellar mass in star-forming spaxels, aiming at understanding the origin of the resolved star-forming main sequence (rSFMS). Our results can be summarized as follows.

1. The three quantities, \sigsfr, \sigh2, and \sigsm, computed at kpc scale, form a 3D linear (in log) relation with dispersion.

2. The 2D projections in each pair of these three parameters show  tight correlations: \sigsfr~ $\propto$ \sigsm$^{1.19}$ (the rSFMS), \sigsfr~$\propto$ \sigh2$^{1.05}$ (the SK relation), and \sigh2~ $\propto$ \sigsm$^{1.10}$ (molecular gas main sequence; MGMS).

3. The power-law exponent (1.05) of the SK relation in our sample is  in good agreement with other recent studies at kpc scales. We also investigate the extended SK law in which a \sigsm~dependence is introduced and we find no significant improvement in the scatter of the relation.

4. The existence of a molecular gas main sequence  implies that either stellar mass dominates the local gravitational potential of the disks or both stars and gas follow the same spatial distributions in response to the gravity set by the underlying total mass.

5. The scatter and correlation analyses suggest that the rSFMS can be naturally explained by the combination of the SK and MGMS relations.

\acknowledgments

We thank the anonymous referee for helpful comments that improved the clarity of this work.
This work is supported by the Academia Sinica under
the Career Development Award CDA-107-M03 and the Ministry of Science \& Technology of Taiwan
under the grant MOST 107-2119-M-001-024 - and 108-2628-M-001 -001 -MY3. R.M. acknowledges ERC Advanced Grant 695671 `QUENCH'.
We thank M. Hani for providing helpful comments to this work. L. Lin and H.-A. Pan thank U. of Victoria for hosting during the visit to work on this project.

Funding for the Sloan Digital Sky Survey IV has been
provided by the Alfred P. Sloan Foundation, the U.S.
Department of Energy Office of Science, and the Participating Institutions. SDSS-IV acknowledges support
and resources from the Center for High-Performance
Computing at the University of Utah. The SDSS web
site is www.sdss.org. SDSS-IV is managed by the Astrophysical Research Consortium for the Participating
Institutions of the SDSS Collaboration including the
Brazilian Participation Group, the Carnegie Institution
for Science, Carnegie Mellon University, the Chilean
Participation Group, the French Participation Group,
Harvard-Smithsonian Center for Astrophysics, Instituto
de Astrof\'isica de Canarias, The Johns Hopkins University, Kavli Institute for the Physics and Mathematics of the Universe (IPMU) / University of Tokyo, Lawrence
Berkeley National Laboratory, Leibniz Institut f\"ur Astrophysik Potsdam (AIP), Max-Planck-Institut f\"ur Astronomie (MPIA Heidelberg), Max-Planck-Institut f\"ur
Astrophysik (MPA Garching), Max-Planck-Institut f\"ur
Extraterrestrische Physik (MPE), National Astronomical Observatory of China, New Mexico State University,
New York University, University of Notre Dame, Observat\'ario Nacional / MCTI, The Ohio State University,
Pennsylvania State University, Shanghai Astronomical
Observatory, United Kingdom Participation Group, Universidad Nacional Aut\'onoma de M\'exico, University of
Arizona, University of Colorado Boulder, University of
Oxford, University of Portsmouth, University of Utah,
University of Virginia, University of Washington, University of Wisconsin, Vanderbilt University, and Yale University.

\end{document}